\documentclass{ws-procs9x6}

\begin{document}

\title{Confining Strings and High-Energy Reactions}

\author{FRANK DANIEL STEFFEN}
\address{
Institute for Theoretical Physics, Bern University\\
Sidlerstrasse~5, CH-3012 Bern, SWITZERLAND\\
E-mail: steffen@itp.unibe.ch}



\maketitle

\abstracts{We present the loop-loop correlation model that allows us
  to compute confining flux tubes and to examine their effects and
  manifestations in high-energy reactions.}

The loop-loop correlation model (LLCM) combines perturbative gluon
exchange with the non-perturbative stochastic vacuum
model.\cite{Shoshi:2002in+X} This combination describes the static
quark-antiquark potential with color Coulomb behavior for small and
confining linear rise for large source separations in agreement with
lattice results. Computing the chromo-electric fields of a static
color dipole in the fundamental and adjoint representation of
$SU(N_c)$, confining flux tubes are obtained for large dipoles that
show exact Casimir scaling.\cite{Shoshi:2002rd} Based on analytic
continuation, the Euclidean LLCM provides the $S$-matrix element
$S_\mathrm{DD}$ for high-energy dipole-dipole
scattering.\cite{Shoshi:2002rd} For reactions involving large dipoles,
we find clear manifestations of the confining strings. Convoluting
$S_\mathrm{DD}$ with appropriate wave functions to describe
dipole-hadron scattering, the confining string in the dipole can be
decomposed into stringless dipoles with a given dipole number density.
This allows us to calculate unintegrated gluon distributions of
hadrons and photons from dipole-hadron and dipole-photon cross
sections via $k_\perp$ factorization.\cite{Shoshi:2002fq}

\end{document}